# Advancing Data for Street-level Flood Vulnerability: Extraction of Variables from Google Street View in Quito, Ecuador

**Raychell Velez**
Lehman College
City University of New York
Bronx, NY, USA
raychell.velez@lc.cuny.edu

**Diana Calderon**
Lehman College
City University of New York
Bronx, NY, USA
diana.calderon@lc.cuny.edu

**Lauren Carey**
Lehman College
City University of New York
Bronx, NY, USA
lauren.carey@lc.cuny.edu

**Christopher Aime**
Lehman College
City University of New York
Bronx, NY, USA
christopher.aime@yorkmail.cuny.edu

**Carolynne Hultquist**
Center for International Earth Science Information Network (CIESIN)
Palisades, NY, USA
c.hultquist@columbia.edu

**Greg Yetman**
Center for International Earth Science Information Network (CIESIN)
Palisades, NY, USA
gyetman@ciesin.columbia.edu

**Andrew Kruczkiewicz**
International Research Institute for Climate and Society (IRI)
Palisades, NY, USA
andrewk@iri.columbia.edu

**Yuri Gorokhovich**
Lehman College
City University of New York
Bronx, NY, USA
yuri.gorokhovich@lehman.cuny.edu

**Robert S. Chen**
Center for International Earth Science Information Network (CIESIN)
Palisades, NY, USA
bchen@ciesin.columbia.edu

## ABSTRACT

Data relevant to flood vulnerability is minimal and infrequently collected, if at all, for much of the world. This makes it difficult to highlight areas for humanitarian aid, monitor changes, and support communities in need. It would be time consuming and resource intensive to do an exhaustive study for multiple flood relevant vulnerability variables using a field survey. We use a mixed methods approach to develop a survey on variables of interest and utilize an open-source crowdsourcing technique to remotely collect data with a human-machine interface using high-resolution satellite images and Google Street View. This paper focuses on Quito, Ecuador as a case study, but the methodology can be quickly replicated to produce labelled training data in other areas. The overall project goal is to build training datasets that in the future will allow us to automate the mapping of flood vulnerability for urban areas in geographic regions.

## 1. INTRODUCTION

Flood hazards occur at the intersection of socio-economic and hydrologic factors. They are specifically exacerbated by urban environments where limited surface water infiltration, artificial "urbansheds" and poorly maintained drainage systems produce high flood risk [7, 26]. Methods of flood vulnerability assessment include considering the relationship of features using regression [4], multicriteria GIS methods [29], integrated vulnerability assessment [20], multiple flood vulnerability assessment [38]. Huang et al. (2012) [18] described three main groups of assessment: disaster loss, vulnerability indices, and vulnerability curves as well as extending a data envelopment analysis (DEA) approach weighting vulnerability categories.

Regardless of specific analytical tools, all the above methods are heavily dependent on data availability, accessibility, and quality, which is often poor in highly vulnerable areas. Moreover, data can be incomplete and/or inconsistent [8] or not available at all. Therefore, alternative crowdsourcing methods can be used to provide fast, easy, and cheap data collection and "data driven" analytical tools to expand training data that can be used in machine learning methods. The presented study might be one of the first to document a method to bridge a mechanical turking method with the development of training data for machine learning based flood vulnerability analysis.

We select the city of Quito, Ecuador as a case study to test out our methodology. Ecuador is a country with diverse terrain with coastal, mountain, and rainforest regions. Ecuador suffers from severe flood events after rainfall. Heavy rainfall events cause rivers to overflow into the built environment, drainage systems to collapse, as well as severe damage to civil infrastructure, housing, agriculture, health impacts, communication, and other systems. Quito is an urban environment with a population of almost 2 million in a valley surrounded by mountainous terrain with exposure to flooding [37]. The concentration and growth of population in urban areas increases the level of exposure to adverse natural events. According to a World Bank report in 2012, approximately 9 million inhabitants of Ecuador lived in urban areas with 96% of these populations living in coastal and mountainous regions, which are also areas where natural hazards occur [36].

In this study, we design a survey and use a mechanical turking approach to capture street-level variables that are associated with flood vulnerability from Google Street View (GSV) and satellite remote sensing imagery [10]. A team of researchers with GIS experience were tasked with selecting appropriate variables to collect, building a consistent dataset by development of a standardized reference for variables, and conducting the data collection through the local Mechanical Turk (MTurk) interface. The constructed dataset will provide the training data for a machine learning model to estimate flood vulnerability in urban areas.

## 2. METHODS

GSV is a feature within Google mapping products which allows users to interact with panoramic views of the Earth's surface [13]. Google collects street images using panoramic cameras, and in addition, displays images that are contributed by the public. Initially released in May 2007 for five U.S. cities, GSV included Quito in November 2015 [9, 19]. By 2019, GSV had reached more than 10 million miles of global street coverage [12, 25]. We utilize a free, open-source local computer version of Mechanical Turk (MTurk) [5] as a platform to collect variables by expert GIS researchers on our team. An interface using the MTurk environment was implemented that required two primary inputs: 1) a survey document that includes our variables of interest for collection, and 2) geographic coordinates associated with remote sensing imagery and GSV.

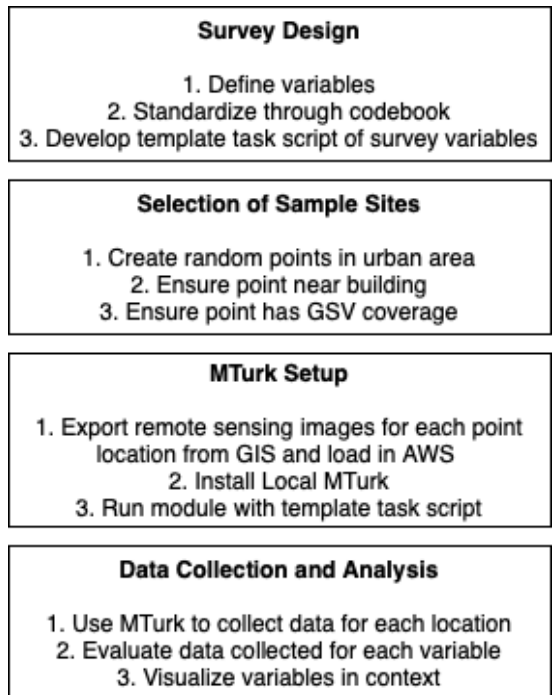

**Figure 1: Flowchart of the methodology from survey design and site selection to the MTurk setup for data collection and the analysis of variables.**

### 2.1 Survey Design

We employed a mixed methods approach and drew on qualitative research to design a survey instrument to collect consistent flood vulnerability variables that could be pulled into MTurk [30]. Through a literature review of flood risk, we selected variables that are critical factors in flood risk assessment, particularly those occurring in urban settings [11, 27, 31, 40]. While the final variables are not exhaustive of all flood-risk factors, we determined that the following list of variables can be collected using GSV imagery. Furthermore, some of these variables had already been collected from GSV in prior research [2, 3, 6, 23]. List of variables:

Number of floors [11, 23]

Sill height [24]

Roof type [24]

Building typology (e.g., residential) [2]

Street material (e.g., cobble, gravel) [11]

Structure attached to adjacent building [3]

Building material [6]

Overall building condition [11]

Occupancy status

Street slope [11]

Early on we realized that our approach requires the development of a standardized way to define variables to ensure consistency in data collection during the turking process. The flood vulnerability variables are classified in a codebook consisting of a written definition and reference images. The authors select the GSV reference images for each variable from several global case studies in order to create a codebook that is globally relevant for all regions. Definitions are compiled from common construction industry descriptions, then discussed and agreed upon among the nine authors. A portion of this codebook is shown in Table 1.

| |
|---|
| Variable 1: Building conditions which measured the functionality of the structure. |
| (1) Very poor: Requires major restoration with possible need to overhaul building subsystems. The approximate restoration cost is 45–60% of building replacement value [32]. |
| (2) Poor: Requires significant updating or restoration. The approximate restoration cost is 30–45% of building replacement value. The physical conditions adversely affect building operations [32]. |
| (3) Fair: Requires updating or restoration. The approximate restoration cost is 15–30% of building replacement value. The physical conditions may influence building operations [32]. |
| (4) Good with minor defects: The approximate restoration cost is 5-15% of building replacement value. |
| (5) Very Good: The approximate restoration cost is less than 5% of building replacement value [32]. |
| Variable 2: Type of roofing material which is sometimes related to socioeconomic status. |
| (1) Corrugated metal: Galvanized sheet iron or sheet steel shaped into straight parallel regular and equally curved ridges and hollows [22]. |
| (2) Tile: Roof tiles are often made from local materials such as terracotta or slate. Modern materials such as concrete, metal and plastic are common, and some clay tiles have a waterproof glaze [34]. |
| (3) Thatched, or palm leaves: Roof with dry vegetation such as straw, water reed, sedge, rushes, heather, or palm branches [35]. |
| (4) Gravel: layer added to the "build-up roof." A layer of gravel, or small stones, is applied on top of the final coating of asphalt to protect the roof from the elements, including ultraviolet (UV) rays and hail. The gravel is embedded into the topcoat of asphalt, which helps the gravel to stay in place [28]. |

**Table 1: Codebook definitions of a few select variables collected through the turking process**

We develop a simple HTML script to collect these variables through the mechanical turk process. This script is essential for the survey to be used in MTurk and to incorporate the necessary images for evaluation. The script can be easily modified when adjustments to the survey are needed. This script is distributed to all the participants to run the modules on their own local computers using the location of the sample sites.

## 2.2 Selection of Sample Sites

We generate 500 random sample points inside the district boundaries of Quito in ArcGIS. These points are exported as a delimited text file and distributed between four GIS users to ensure each point is near a building and has GSV coverage in order to complete the turking process. A basemap remote sensing layer is used to observe if the points are within view of building locations. If a point does not fall near a building location, it is moved to be near the closest building. We also verify that the points are located close to buildings with GSV coverage. If a building layer is available, then this process can be expedited by constraining the random point generation to within a buffer of the building layer.

## 2.3 MTurk Setup

Turking is a form of crowdsourcing which breaks down tasks into simpler components with easier decision processes then distributes these focused tasks to many individuals. The turking platform facilitates a convenient approach to respond to the tasks and record the inputs [1]. Turking approaches have had a wide range of applications, from assessing comprehensibility of medical pictograms [39] to studying sidewalk accessibility problems in concert with GSV [16]. Turking approaches can also be informed by traditional survey methods to ensure that appropriate questions are being asked.

MTurk requires input of the survey template we designed through the literature and the 500 geographic coordinates of our sampled points. In addition, we export satellite remote sensing imagery for each point from ArcGIS using Data Driven Pages to capture each footprint at an appropriate scale to view the building. These images are loaded into Amazon Web Services (AWS) as data to be accessed by link from MTurk.

We installed a local version of MTurk by Danvk (2018) [5] which is available through GitHub and enables the automated turking setup on your own computer. We used this environment to have our team of GIS experts provide the inputs on this repetitive task in an easy-to-use interface. We decided to use this implementation at this stage to strive to get consistent results from trained users in our group instead of relying on random and potential non-expert Turkers from the crowd. This was important for our group to continue the development of the survey questions through these experiences and ensure that the results are reasonable. We believe that the project could be expanded to the online MTurk platform once the project is fully developed and if user training is provided on using the codebook for standardization.

Finally, we ran the local MTurk module with the template task script for our locations of interest. As shown in Figure 2, the survey document interface uses the HTML template task script that pulls from the list of sample sites and includes both the AWS link to the footprint image and a link to GSV for the turkers to review the area further.

**Figure 2: Turking Survey Interface**

## 2.4 Data Collection and Analysis

The GIS users from our team collected data at each geographic coordinate sample site point by examining the corresponding GSV image and recording variable characteristics using our survey. Some responses allowed for a free form "Other" response if the selection was not available. In some cases, users found that open responses were common enough that the survey template was updated to incorporate these changes. We evaluated the data collected for each variable through basic exploratory analysis. This step was primarily to ensure the reasonableness of the variables and do checks in a few spots in GIS. Finally, we visualized the collected data in ArcGIS to identify initial spatial patterns of the flood vulnerability variables (e.g., number of drains in Figure 4).

## 3. RESULTS

The mechanical turking approach presented an alternative to the utilization of on the ground field surveys. Data about variables were collected at the sample points in Quito, Ecuador. Google Street View (GSV) within the MTurk interface was the vehicle used to virtually navigate the study area in a short amount of time. We collected data for 458 points in Quito and were unable to collect data for 42 points, either because the points were too close together or GSV coverage was not available. In the present case study for Quito, the average time for the turking process was 3 - 5 minutes per point with each location having 12 variables to collect.

The variables were observed through two remote geospatial technologies used within the MTurk interface. On the top half of Figure 3 you can observe an overview of the surroundings by a satellite remote sensing view from above. This image can help to add context about the building location and environment. The bottom half of Figure 3 has a GSV panoramic image that can be used to move along the street just like in a Google Map environment. The associated variables were collected from simultaneously viewing both the satellite and street-level views seen in Figure 3. These observations helped the collection of data about the roof type, street material, land use, and slope. Navigating the street with the coverage provided by GSV particularly allowed for the collection of information about the sill height, slope, number of drains, building condition and material, and number of stories.

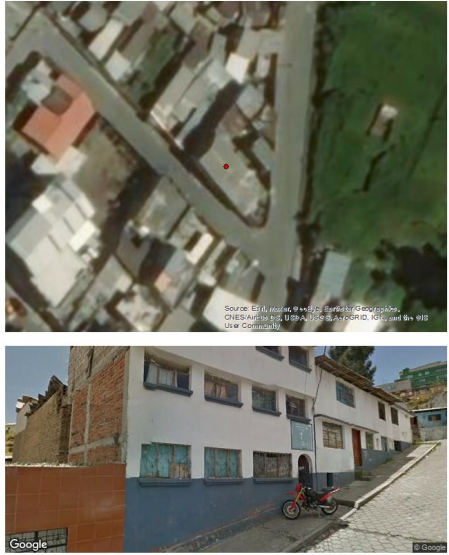

**Figure 3: Example of building point with corresponding remote sensing image and GSV**

The data collected about the variables were visualized in ArcMap as seen in Figures 4 and 5. This is just one example to demonstrate the data that were collected. The visualization illustrates the ability to quickly capture a characteristic of the city relevant to flooding from anywhere GSV is available.

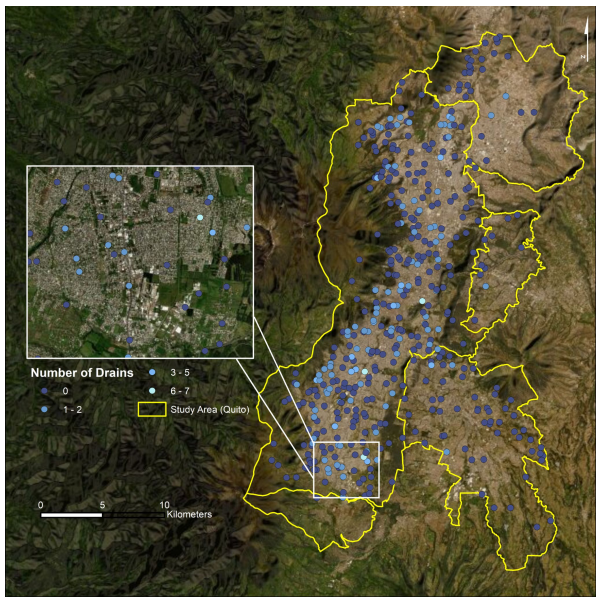

**Figure 4: Map of variable collected on number of drains visible at specific point locations.**

The map depicted in Figure 4 shows the number of drains found at each point surveyed through the mechanical turking process. The darkest blue points show locations of buildings where drains were not visible at the street level, suggesting a lack of drains along the periphery of the city. In particular, the southeast part of Quito shows a cluster of zero drains. This could indicate a higher grade of vulnerability to flooding in that area. The light blue colored points represent the location of buildings where drains were visible; most of these occur close to the center of the city of Quito.

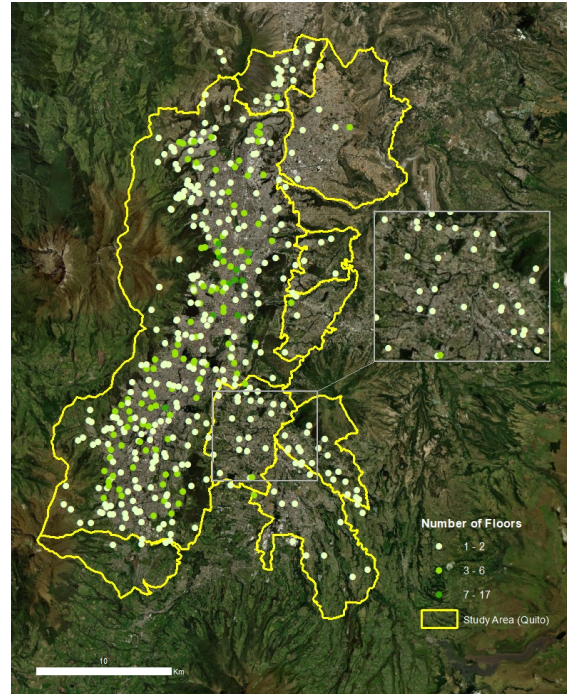

**Figure 5: Map of observations collected on the number of floors at point locations.**

Figure 5 shows the spatial distribution of the sampled buildings, with darker green representing buildings with more floors (stories) and lighter green indicating those with fewer floors. The periphery of the city, in particular the southeastern part of Quito in the inset, shows a cluster of low-height buildings but almost no taller buildings. The number of floors has been included in building vulnerability indices for floods [11], so the spatial distribution of this information across Quito is valuable for flood risk assessment.

The data we collected are relevant for future analysis of flood risk to buildings in Quito. For example, the number of drains (Fig. 4) relates to the status of existing infrastructure to counter floods surrounding a building. Cursory observation of Figure 4 demonstrates that drain coverage varies greatly in Quito - some areas lack drains while others have several. This variability in infrastructure is relevant for decision-makers and emergency management, allowing them to take into account building and street-level variables for spatial flood risk assessment. Furthermore, the ease of data collection without the need for in situ observation potentially reduces the financial and capacity burden for emergency management.

## 4. DISCUSSION

We see certain advantages of the street-level data collection and modeling for humanitarian mapping as opposed to traditional hydrologic/hydraulic engineering approaches. Typical hydrologic models for urban flooding provide a spatial map of runoff depth and discharge that helps to plan engineering intervention to reduce flood risk. The advantage of collecting crowdsourced street-level data is that it directly reflects the socioeconomic status of the area and physical characteristics (e.g., roof type, building material, building condition, sill height, presence of drains, street inclination). For example, high poverty areas might not have high sills at the doorsteps and are likely to be located on steep slopes at the periphery of Quito with poor construction.

Integration of hydrologic models and social vulnerability characteristics could provide an ideal framework for flood vulnerability mapping. The hydrologic/hydraulic process requires significant resources to collect all the necessary data along with consequent model calibration and verification. From this point of view, the process for collecting data for vulnerability is perhaps faster to implement for humanitarian purposes and feasible to accomplish remotely. The data could then be used alongside flood observations and models, or other hazard layers.

Unfortunately, there is not comprehensive GSV coverage of streets globally as many areas lack collection, particularly in the developing world and in rural areas, a concern is that many of these areas are those in which flood risk assessment and flood anticipatory action programs are most needed [21]. It is noted that the methods presented here will have limitations on where they will be applicable, and further, we note that data quality using GSV may vary even in areas where available, therefore careful consideration must be taken to ensure appropriate adjustments for standardized collection are made. These adjustments are done by moving the previously generated points to areas with available GSV. The lack of images in certain places may be a result of a sampling bias in collection despite the randomized method and could cause an unintended misrepresentation of study areas.

While GSV coverage is extensive across Quito, it is not available for every building. The team often encountered buildings without corresponding GSV imagery which resulted in the final collection of variables for 458 points out of 500. Some variables were more difficult to collect than others. For instance, roof material type was not always visible from GSV. Assessing street slope from GSV also proved challenging, particularly when distinguishing between low and medium slopes. In terms of the consistency of variable collection, a level of subjective interpretation during turking is unavoidable even after standardization of variable descriptions.

We also encountered limitations in the view from the street due to physical barriers. In Quito, many residential and commercial buildings have high walls around them resulting in the inability to view the complete building via GSV. This prevented an accurate assessment of the physical structure of the buildings. During our background research we noticed mention of geotagging inaccuracies when attempting to locate structures in GSV based on addresses [6]. Our study, which used coordinates rather than street addresses to locate buildings, also encountered this issue. However, while Diakakis et al. (2017) [6] were able to use a building address number to ensure the correct structure was identified, we had to rely on comparison with the footprint image. These issues should be considered when determining how to use GSV derived data.

Despite the current limitations on GSV coverage, we believe that there is the potential to continue to expand this approach to other urban areas. With the addition of new resources and ways of collecting, Google is expanding its coverage every year [33]. Google Street View trekkers is a program that provides camera equipment to volunteer professional photographers and other individuals to collect GSV images [15, 17]. In addition, one can hire trusted professionals to produce Street View in particular areas [14]. There is also a Street View app in which anyone can collect GSV through mobile devices and upload a path [33]. These alternative ways of inputting data can reduce some of the barriers of coverage, provide the ability to task coverage in areas of interest, and open more areas for assessment.

## 5. CONCLUSIONS

Natural disasters disproportionately impact vulnerable populations who are often not highlighted on the map. We present a crowdsourcing method using a human-machine interface to map characteristics of the built infrastructure that also provide socioeconomic context relevant to flood vulnerability in urban areas. This work shows that we can use remote collection techniques to capture information on vulnerability which is otherwise unavailable, infrequent, or would require time-consuming and expensive field studies. We learned that we can quickly capture flood vulnerability variables on hundreds of buildings in a few hours based on visual inspection.

While this paper focuses on Quito, Ecuador as a case study, the methodology can be quickly replicated to produce labelled training data in other areas. The long-term goal of this project is to build training datasets for urban areas for global regions that will allow us to automate the mapping of flood vulnerability in the future. To this end, future work involves the use of an inter-rater validation technique to evaluate the agreement in the turking results of each of the participants for a selection of common cases in Quito, Ecuador. Our approach needs to show consistency amongst users which is why the standardized codebook was developed to encourage a common understanding. We also need to ensure that the data collected and the trends within cities are accurate. One step in this direction is the mapping of the variables collected along with other datasets on the same topics, such as the slope variable mapped with an elevation layer.

Further work comparing data collected through our MTurk approach with census microdata at the building level through field studies would complement this initial exercise. However, the availability of detailed microdata is limited and is not available for Quito, Ecuador. We are developing a separate case study in a city in Colombia that has official field study microdata from the national census to further validate the collection of the labelled data. In addition, our method could be adapted for participatory GIS approaches with community partners to target specific areas of interest for evaluation of the outputs based on local knowledge.

This method for collection along with the upcoming evaluation steps and the expansion of this project lead the way for remote collection of street-level variables to contribute valuable information on vulnerability for humanitarian action. High resolution information on factors affecting flood vulnerability can be an invaluable resource for all stages of the disaster management cycle. The ability to collect data quickly and remotely on street-level conditions could be a game changer for decision makers. We believe that the use of novel geospatial technologies is an emerging research area that will have significant implications for humanitarian policy as high resolution data become more timely and accessible.

## 6. ACKNOWLEDGMENTS

We acknowledge funding from NASA Goddard Space Flight Center (GSFC) contract 80GSFC18C0111, Continued Development and Operation of Socioeconomic Data and Applications Center as a Distributed Archive Center, led by the Center for International Earth Science Information Network (CIESIN) at Columbia University, and NASA contract 80NSSC18K0342.

## About the authors:

**Raychell Velez** graduated from the Master of Science in Geographic Information Sciences (GISc) at the City University of New York (CUNY), Lehman College. She works as a high school science research teacher and college counselor. Website: https://raychellvelez.com Email: raychell.velez@lc.cuny.edu

**Diana Calderon** is a M.S. Geographic Information Sciences (GISc) student at the City University of New York (CUNY), Lehman College. She is an intern at the Census Bureau and a research assistant for CIESIN. Email: diana.calderon@lc.cuny.edu

**Lauren Carey** is a M.S. Geographic Information Sciences (GISc) student at the City University of New York (CUNY), Lehman College and a research assistant for CIESIN. Email: lauren.carey@lc.cuny.edu

**Christopher Aime** is a M.S. Geographic Information Sciences (GISc) student at the City University of New York (CUNY), Lehman College and a research assistant for CIESIN. His interests are in flood risk, wildlife conservation, and remote sensing. Email: christopher.aime@yorkmail.cuny.edu

**Carolynne Hultquist** is a Postdoctoral Research Scientist at CIESIN. She specializes in fusing sources of geographic information to better understand complex environments. Her research focus is on the development of computational methods for spatio-temporal analysis and modeling of environmental hazards. Email: c.hultquist@columbia.edu

**Greg Yetman** is associate director of the Geospatial Applications Division at CIESIN. He is a geographer specializing in the application of geographic information system (GIS) technologies in applied research fields, including population geography, natural disasters, and environmental assessment. Email: gyetman@ciesin.columbia.edu

**Andrew Kruczkiewicz** is a Senior Researcher at Columbia University, based at the International Research Institute for Climate and Society. He is faculty within the Climate and Society graduate program of Columbia's Climate School. Also, Andrew is science advisor to the Red Cross Red Crescent Climate Centre. Email: andrewk@iri.columbia.edu

**Yuri Gorokhovich** is a physical geologist who combines geologic and geographic methods, including GIS for the assessment, modeling, and mapping of current and historical natural hazards and disasters. He is an Associate Professor at the City University of New York (CUNY), Lehman College. Email: yuri.gorokhovich@lehman.cuny.edu

**Robert S. Chen** is the director of CIESIN, a unit of the Earth Institute at Columbia University. He manages the NASA Socioeconomic Data and Applications Center (SEDAC) and co-leads the Human Planet Initiative of the Group on Earth Observations. He also serves as a Councilor of the American Geographical Society. Email: bchen@ciesin.columbia.edu